# Non-Uniform Quantum Well and Barrier Thickness Engineering for Robust Ultra-High TER and Low RA Ferroelectric Tunnel Junctions

Balram Khattar, Adarsh Tripathi, Manish Anand, *IEEE*, and Abhishek Sharma, *Member, IEEE*

***Abstract*—** **$HfO_2$-based ferroelectric tunnel junctions (FTJs) are promising candidates for scalable non-volatile memory, but simultaneously achieving a high tunneling electro-resistance ratio (TER) and a low resistance-area (RA) product remains challenging. To address this challenge, this work introduces non-uniform quantum well (QW) and barrier thickness engineering in $HfO_2$-based multi-QW FTJs using a self-consistent Preisach-based ferroelectric (FE) model integrated with the coherent and inelastic non-equilibrium Green's function (NEGF) formalism. The non-uniform well and barrier configuration produces a wide range of FTJ design landscapes due to closely spaced, broad resonant states in the low resistance state (LRS) and a larger separation in the high resistance state (HRS), resulting in strong polarization-dependent resonant transmission with TER reaching the order of $1 \times 10^8\%$ and an LRS RA product as low as $1\,\Omega \cdot \mathrm{cm}^2$ at read bias and in the presence of scattering. By incorporating self-consistent elastic scattering into the NEGF framework, we also show that elastic scattering can positively influence the TER and RA performance of non-uniform FTJs by progressively increasing the overlap of the closely spaced resonances. Overall, the results establish non-uniform QW and barrier thickness as an effective and robust design parameter for controlling resonant-state alignment and achieving a favorable combination of high TER and low RA in $HfO_2$-based multi-QW FTJs.**

## I. Introduction

$HfO_2$-based ferroelectric tunnel junctions (FTJs) have attracted considerable interest for nano-scale non-volatile memory (NVM) owing to their CMOS compatibility, low switching energy, fast polarization reversal, and robust ferroelectricity at reduced thicknesses [1]–[3]. For practical FTJ memory, a high tunneling electro-resistance ratio, $\mathrm{TER}(\%) = [\mathrm{J_{LRS}}/\mathrm{J_{HRS}} - 1] \times 100$, is required for reliable state distinction, where $\mathrm{J_{LRS}}$ and $\mathrm{J_{HRS}}$ denote the current densities of the LRS and HRS, respectively, while a low RA product is important for high read current, compact device dimensions, and high packing density [4], [5]. Achieving both characteristics remains challenging because a high TER can be associated with a reduced LRS current density, resulting in a higher RA product [6].

Resonant tunneling (RT) can enhance electron transmission when quantum-confined states align in energy with the transport window [7], [8]. In multiple-QW structures, barrier thickness influences the coupling between neighboring wells through wave-function penetration, thereby affecting resonant transport [9]–[11]. Studies of superlattice structures have further shown that well and barrier dimensions, including non-uniform layer profiles, can be used to tailor the transmission characteristics [12], [13].

In FTJs, RT has been demonstrated in single-QW structures, where electrostatic-potential variations shift the resonant states and produce distinct transmission characteristics between the LRS and HRS [14]–[17]. Previous studies of multi-QW FTJs have investigated the effects of QW number and well thickness on tunneling conductance and TER through RT [18]. More recently, $HfO_2$-based multi-QW FTJs have been studied with different QW numbers and layer thicknesses, demonstrating the role of RT and polarization-dependent transport in TER enhancement [19]. However, systematic studies of non-uniform QW and IL thickness distributions for controlling resonant-state alignment and the resulting TER–RA characteristics in $HfO_2$-based multi-QW FTJs remain limited. QW thickness primarily modifies the quantum-confinement energy and hence the resonant-state energies, whereas IL thickness influences wave-function penetration between neighboring wells and the resulting resonant transmission [9], [20]. These thicknesses therefore provide distinct structural parameters for tuning resonant transport. A systematic investigation of their distributions is needed to clarify their effects on polarization-dependent RT and TER–RA characteristics.

In this work, we systematically investigate thickness-engineered multi-QW $HfO_2$-based FTJs using a self-consistent Preisach-based FE model [21], [22] integrated with the NEGF formalism [23], [24]. FE-defined and IL-defined QW-FTJ architectures with 0–3 QWs are first compared in terms of their TER and RA characteristics to identify the structure for subsequent analysis. The selected double-QW architecture is then investigated using uniform QW thicknesses to establish a reference for subsequent thickness engineering. Non-uniform QW thicknesses are subsequently introduced while maintaining uniform IL thicknesses to tune the relative resonant-state alignment. The selected QW dimensions are then retained

Balram Khattar and Adarsh Tripathi are with the Department of Electrical Engineering, IIT Ropar, Rupnagar, Punjab 140001, India.

Manish Anand is with the Department of Physics, Bihar National College, Patna University, Patna-800004, India.

Abhishek Sharma is with the School of Computing and Electrical Engineering, Indian Institute of Technology Mandi, Mandi 175005, India, India (e-mail: abhi@iitmandi.ac.in).

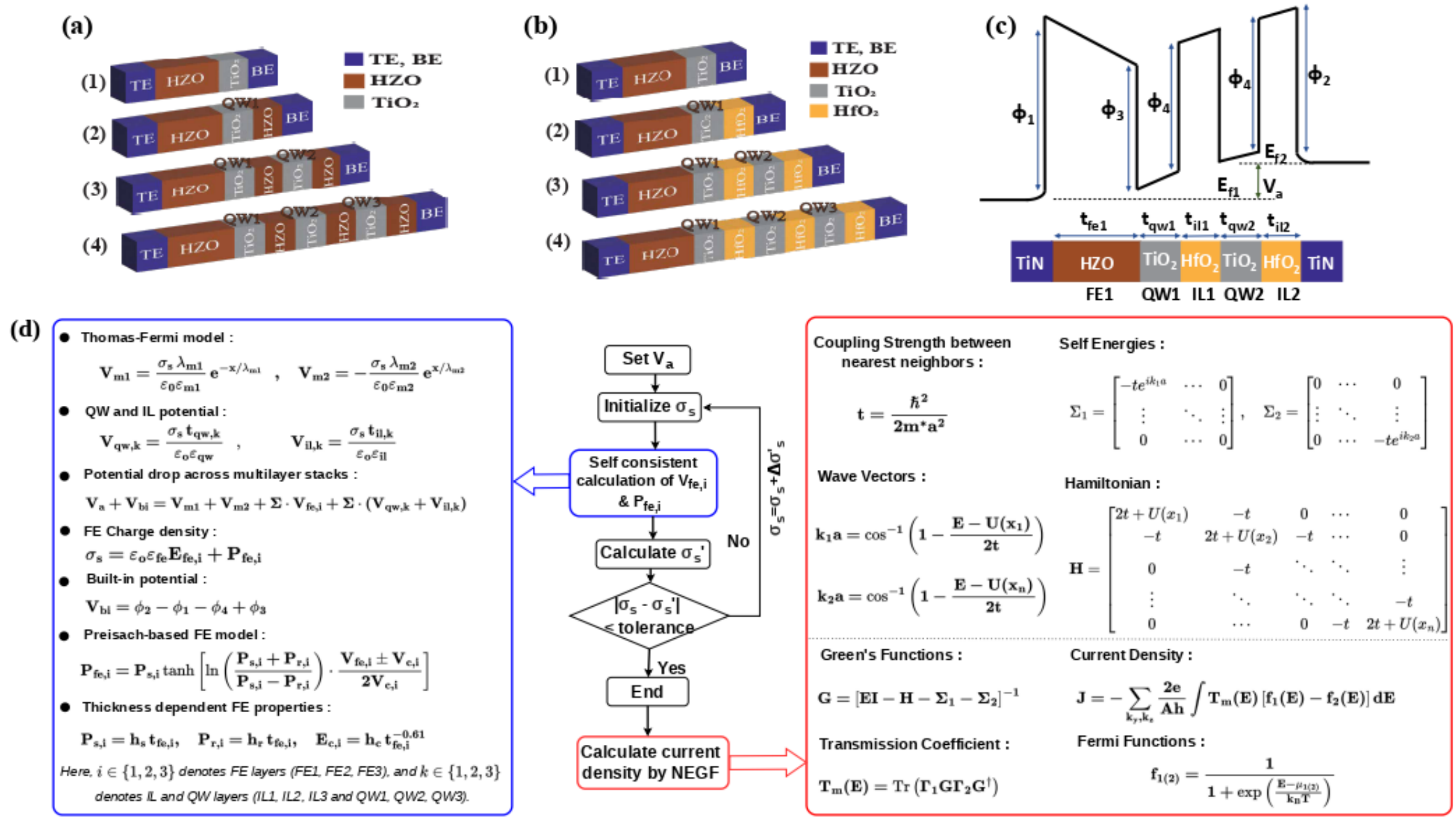

Fig. 1. Schematic illustration of the FTJ device architectures shown for 0 to 3 QWs: (a) FE layer-defined QW-FTJs, where QWs ($\mathbf{TiO_2}$) are formed between FE HZO barriers; (b) IL-defined QW-FTJs (MFIIIIM-type), where alternating layers of $\mathbf{TiO_2}$ and $\mathbf{HfO_2}$ serve as the QWs and tunneling barriers, respectively; (c) Energy-band diagram of the 2-QW FTJ-(b3) under bias voltage ($\mathbf{V_a}$), and (d) Self-consistent simulation framework, using a Preisach-based FE model integrated with the NEGF formalism for FE-layer and IL-defined stack configurations.

while varying the IL thicknesses to provide additional control over resonant transport. The polarization-dependent resonant states and transmission characteristics are analyzed, while self-consistent elastic scattering is examined to assess its effect on the resonant response. This systematic approach clarifies the roles of QW and IL thickness distributions in tailoring resonant transport and provides a basis for designing multi-QW FTJs with high TER and low RA characteristics for practical NVM applications.

## II. Device Structure and Modeling

Fig. 1 illustrates the proposed QW-engineered FTJ architectures. Two distinct architectures are considered with 0–3 QWs, namely, FE layer-defined QW-FTJs, in which $TiO_2$ QWs are formed between ferroelectric HZO barriers, and IL-defined QW-FTJs, in which $TiO_2$ and $HfO_2$ layers form the QWs and tunneling barriers, respectively. The FE layer-defined QW-FTJs in Fig. 1(a) consist of TiN top and bottom electrodes (TE and BE), HZO FE layers, and $TiO_2$ QWs, with the corresponding 0–3 QW configurations shown in Fig. 1(a1)–1(a4). In contrast, the IL-defined QW-FTJs in Fig. 1(b) comprise TiN electrodes, a single HZO FE layer, $TiO_2$ QWs, and $HfO_2$ ILs, with the corresponding 0–3 QW configurations shown in Fig. 1(b1)–1(b4). Fig. 1(c) presents the schematic energy-band diagram of FTJ-(b3), containing two QWs. Under an applied bias ($V_a$), polarization reversal modifies the tunneling potential profile and the alignment of the quantum-confined states relative to the electrode Fermi levels. The resulting resonant transport depends on the FE barrier height, effective FE thickness ($t_{fe1}$), QW widths ($t_{qw1}$ and $t_{qw2}$), and IL thicknesses ($t_{il1}$ and $t_{il2}$). A self-consistent computational framework combining a Preisach-based FE polarization model [21], [22] with the NEGF formalism is employed to capture the coupled FE response and quantum transport [8], [25], as illustrated in Fig. 1(d). For a given applied bias, the FE polarization state is first determined using the Preisach-based model, and the resulting polarization-induced electrostatic potential is incorporated into the NEGF framework to calculate the transmission and current density (J). To assess the effect of scattering on resonant transport, self-consistent elastic scattering is incorporated through additional self-energy and in-scattering terms, which account for momentum and phase relaxation. The scattering potential provides an effective description of disorder-induced scattering associated with mechanisms such as impurities, alloy disorder, and surface roughness. The total retarded self-energy and in-scattering function are given by

$$\Sigma(E) = \Sigma_1(E) + \Sigma_2(E) + \Sigma_0(E), \quad (1)$$

$$\Sigma^{in}(E) = \Gamma_1(E)f_1(E) + \Gamma_2(E)f_2(E) + \Sigma_0^{in}(E), \quad (2)$$

where $\Sigma_{1,2}$ and $\Gamma_{1,2}$ denote the contact self-energies and broadening functions of the TE and BE, respectively. The elastic-scattering contributions are described by

$$\Sigma_0(E) = D \times G^R(E), \qquad \Sigma_0^{in}(E) = D \times G^n(E), \quad (3)$$

where D is the correlation matrix of the random scattering potential, and $\times$ denotes element-by-element multiplication. Here, $G^R$ and $G^n$ represent the retarded Green's function and electron-correlation function, respectively. For local, uncorrelated elastic scattering, the correlation matrix is given by $D_{mn} = \delta_{mn} D_0$, where $D_0$ denotes the dephasing strength. Since the scattering self-energies depend on the Green's functions, the NEGF equations are solved self-consistently for each selected value of $D_0$ to evaluate its effect on resonant transport [26]–[28].

The simulation parameters are listed in Table I. The relative permittivity and effective electron mass of HZO are assumed to be equal to those of $HfO_2$. The TiN electrodes screening lengths $\lambda_{m1}$=$\lambda_{m2}$=0.5 Å [29]. A FE HZO layer with a thickness of 4 nm, having remanent polarization $P_r$=5 $\mu C/cm^2$, saturation polarization $P_s$=5.5 $\mu C/cm^2$, and coercive field $E_c$=1 MV/cm, is used as the reference for determining the thickness-dependent scaling coefficients $h_r$, $h_s$, and $h_c$ [23], [29]. Throughout the study, the FE1 layer at the TE/FE interface is maintained at a fixed thickness of 3 nm, whereas the thicknesses of the remaining FE, QW, and IL layers are varied systematically according to the specific thickness-engineering configuration. The model has been previously calibrated in work [17] against transmission characteristics, consistent with prior studies [16], [30].

TABLE I
MATERIAL PARAMETERS USED IN SIMULATION

| Parameter | HZO | $TiO_2$ | Parameter | TiN |
|---|---|---|---|---|
| Electron Affinity, $\chi$ (eV) | 2.8 [31] | 4.0 [32] | Work Function, $\Phi_m$ (eV) | 4.3 [33] |
| Relative Permittivity, $\varepsilon_r$ | 25 [34] | 48 [35] | Relative Permittivity, $\varepsilon_r$ | 3 [36] |
| Effective Mass, $m^*$ ($m_0$) | 0.14 [37] | 0.5 [38] | Effective Mass, $m^*$ ($m_0$) | 1.5 [39] |

## III. RESULTS AND DISCUSSION

### A. QW-FTJ Architecture and QW-Number Screening

Fig. 2(a) and Fig. 2(b) compare the TER and RA characteristics of the FE layer-defined and IL-defined QW-FTJ architectures with 0–3 QWs, corresponding to the structures shown in Fig. 1(a) and Fig. 1(b), respectively, at a read voltage of $V_r$=12.5 mV. The FE and IL layers, excluding the first FE1 layer (3 nm), are maintained at a thickness of 1 nm, while the QW thickness is fixed at 1.5 nm. The results show that QW incorporation significantly alters the TER and RA characteristics through the formation of quantum-confined resonant states. Increasing the QW number introduces additional resonant states and inter-well coupling, which modify the transmission characteristics and strongly affect the TER and RA. Polarization reversal alters the electrostatic potential profile and shifts the resonant states relative to the electrode Fermi levels, resulting in a distinct transmission contrast between the LRS and HRS. However, additional QWs increase the effective tunneling length and introduce additional barriers, potentially reducing the overall transmission and increasing the RA. Among the investigated configurations, the double-QW IL-defined FTJ-(b3) structure is selected as the reference structure for subsequent thickness engineering based on its favorable TER–RA combination.

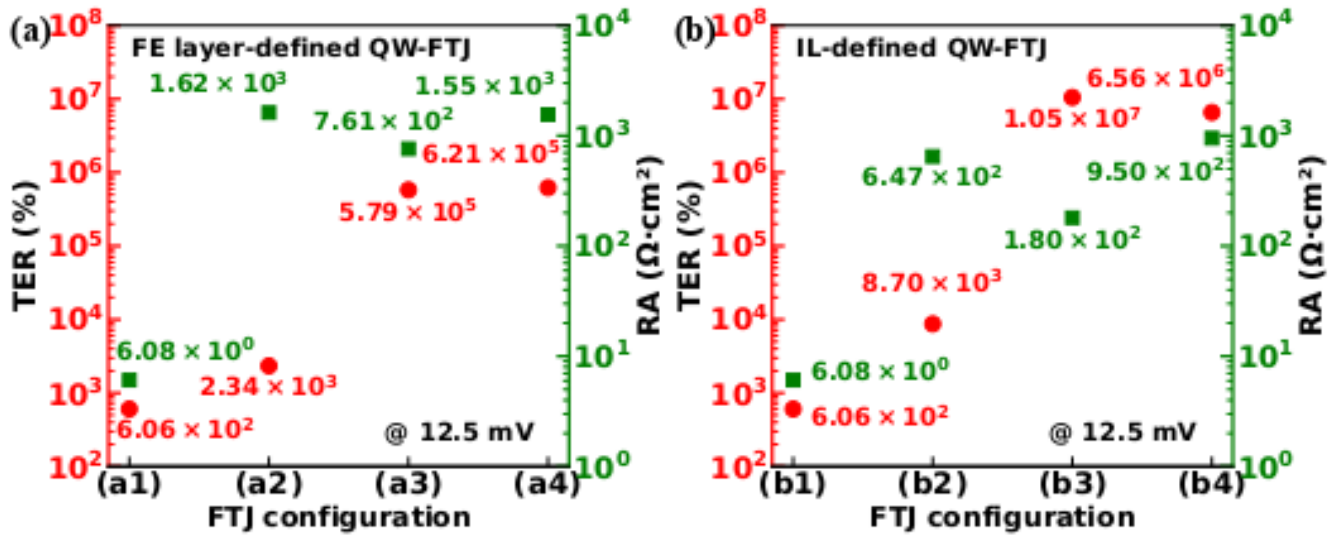

Fig. 2. TER and RA for different QW-FTJ structures evaluated at $V_r$=12.5 mV. (a) and (b) correspond to FE layer-defined and IL-defined QW-FTJs, respectively. Configurations (a1)-(a4) and (b1)-(b4) denote devices containing 0-3 QWs.

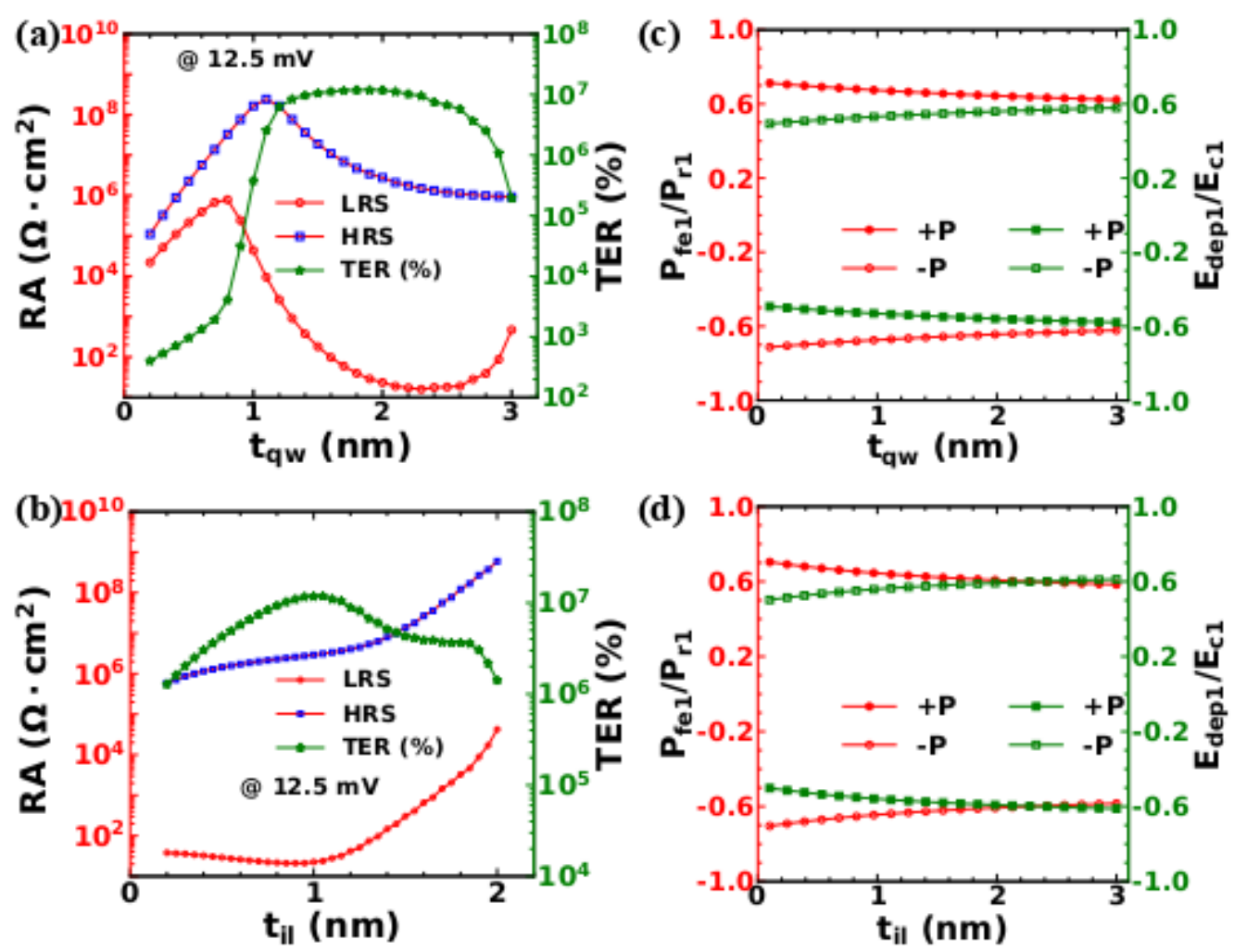

Fig. 3. Variation of the RA and TER, evaluated at $V_r$=12.5 mV, as a function of (a) the equal QW thickness ($t_{qw1}$=$t_{qw2}$=$t_{qw}$) with fixed IL thicknesses of $t_{il1}$=$t_{il2}$=1 nm, and (b) the equal IL thickness ($t_{il1}$=$t_{il2}$=$t_{il}$) with fixed QW thicknesses of $t_{qw1}$=$t_{qw2}$=2 nm for the FTJ-(b3) stack. The corresponding $P_{fe1}/P_{r1}$ and $E_{dep1}/E_{c1}$, evaluated at 0V, are shown in (c) and (d), respectively.

### B. Uniform Thickness Engineering

Having identified the double-QW IL-defined FTJ-(b3) structure as the reference architecture, the effect of uniform QW and IL thicknesses is first investigated to establish a baseline for subsequent thickness non-uniformity engineering. Fig. 3(a) shows the TER and RA as functions of the QW thickness, $t_{qw}$, with equal QW thicknesses ($t_{qw1}$=$t_{qw2}$=$t_{qw}$) and fixed IL thicknesses of $t_{il1}$=$t_{il2}$=$t_{il}$=1 nm. As $t_{qw}$ increases, the TER initially increases while the RA decreases, with the TER reaching a maximum at $t_{qw}$=2 nm. Further increasing $t_{qw}$ reverses these trends, resulting in a decrease in TER and an increase in RA. This behavior is mainly attributed to the thickness-dependent quantum confinement, which shifts the resonant-state energies and changes their alignment within the transport window. The effect of uniform IL thickness is then examined by fixing the QW thicknesses at $t_{qw1}$=$t_{qw2}$=2 nm and varying $t_{il1}$=$t_{il2}$=$t_{il}$. As shown in Fig. 3(b), the RA increases with increasing $t_{il}$, reflecting the reduced tunneling probability through the thicker IL barriers. In contrast, the TER initially increases, gradually saturates, and then decreases slightly. This behavior results from the competing effects of barrier transparency and inter-well coupling on the polarization-dependent resonant transmission. The corresponding normalized polarization and depolarization field, shown in Fig. 3(c) and Fig. 3(d), respectively, show only small variations with increasing QW and IL thicknesses. The normalized polarization, $\mathrm{P_{fe1}/P_{r1}}$, decreases slightly, whereas the normalized depolarization field, $\mathrm{E_{dep1}/E_{c1}}$, increases slightly. These relatively small variations indicate that the pronounced changes in TER and RA are mainly associated with thickness-dependent resonant transport, with only a minor contribution from changes in the FE polarization state.

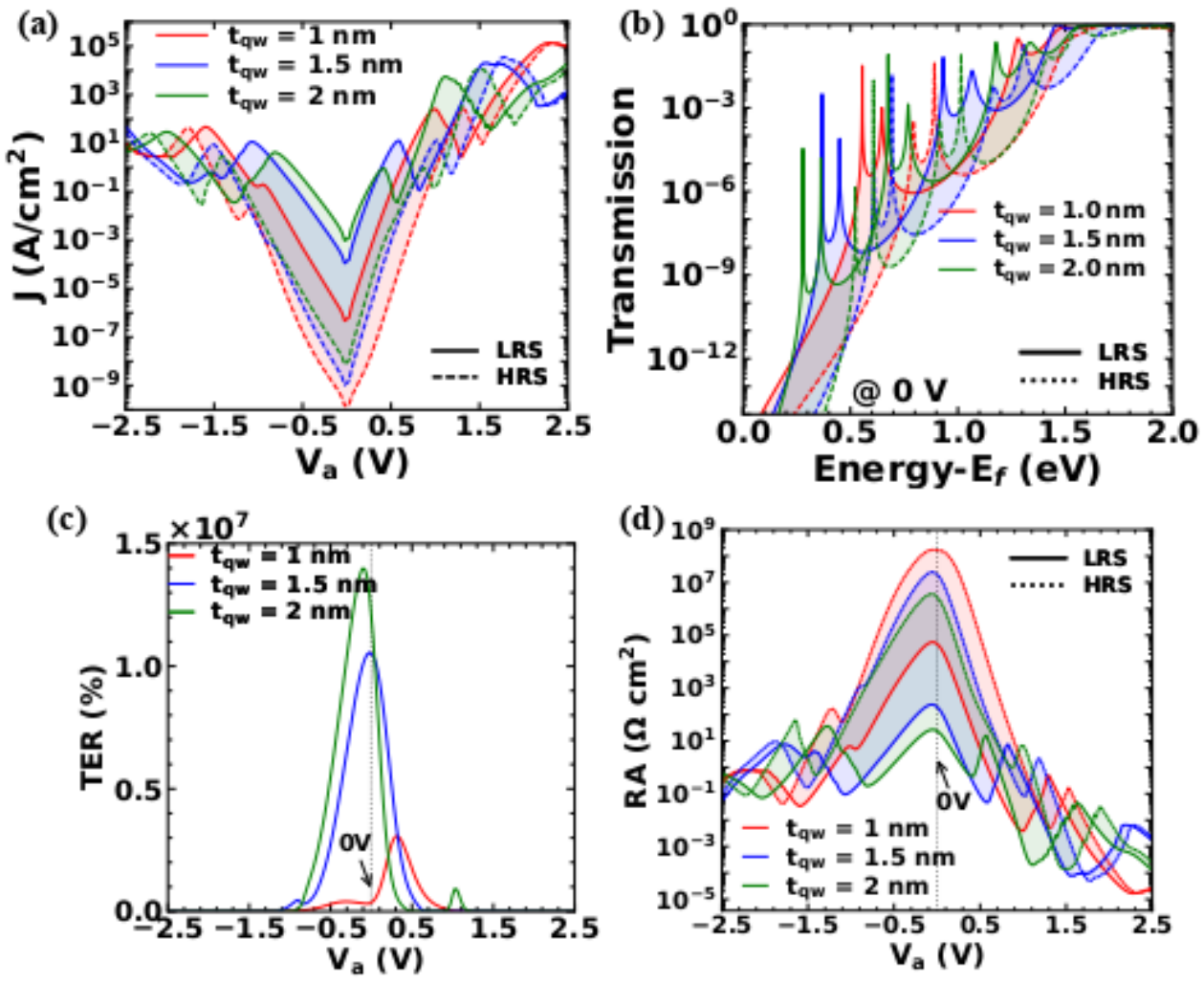

Fig. 4. Transport characteristics of the FTJ-(b3) structure for uniform QW thicknesses of $\mathbf{t_{qw}}$=1, 1.5, and 2 nm with $\mathbf{t_{il}}$=1 nm: (a) J-$\mathbf{V_a}$ characteristics, (b) transmission spectra, (c) TER, and (d) RA as functions of applied voltage ($\mathbf{V_a}$).

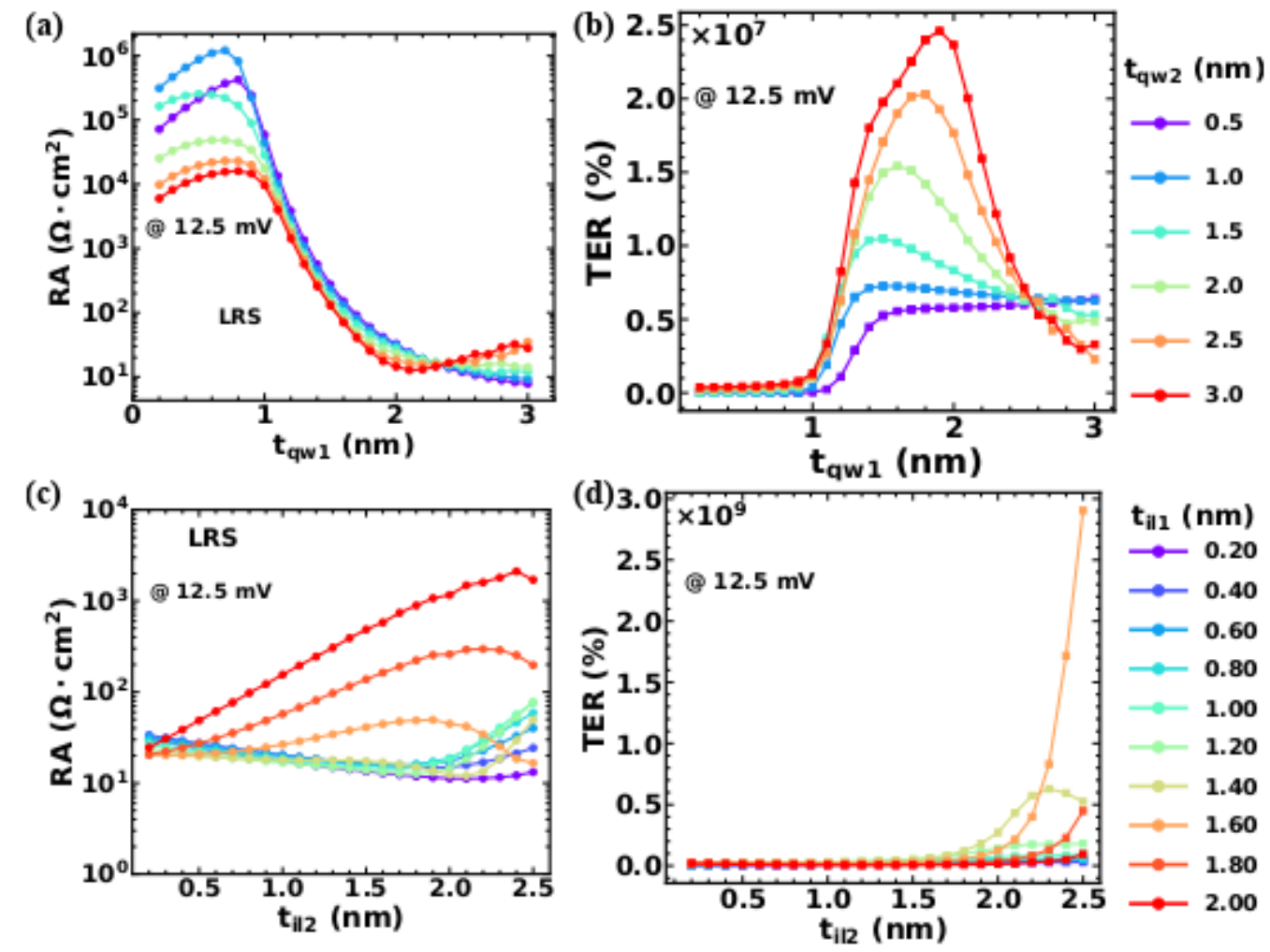

Fig. 5. Thickness-dependent transport characteristics of the FTJ-(b3) structure evaluated at $\mathbf{V_r}$=12.5 mV. (a) RA and (b) TER as functions of the QW thicknesses, $\mathbf{t_{qw1}}$ and $\mathbf{t_{qw2}}$, with fixed IL thicknesses of $\mathbf{t_{il1}}$=$\mathbf{t_{il2}}$=1 nm. (c) RA and (d) TER as functions of the IL thicknesses, $\mathbf{t_{il1}}$ and $\mathbf{t_{il2}}$, with fixed QW thicknesses of $\mathbf{t_{qw1}}$=1.9 nm and $\mathbf{t_{qw2}}$=3 nm.

To further examine the voltage-dependent transport associated with QW thickness, the uniform FTJ-(b3) structure is investigated for $\mathrm{t_{qw}}$=1, 1.5, and 2 nm with $\mathrm{t_{il}}$=1 nm, as shown in Fig. 4. Fig. 4(a) presents the J-$\mathrm{V_a}$ characteristics, showing distinct LRS and HRS currents, particularly in the low-bias region. The corresponding transmission spectra in Fig. 4(b) exhibit multiple resonant peaks whose positions shift with QW thickness due to changes in the quantum-confinement energy. For $\mathrm{t_{qw}}$=2 nm, a pronounced resonant transmission response occurs in the low-bias region, resulting in a relatively large LRS–HRS current difference and consequently a higher TER. This thickness dependence is further reflected in the TER characteristics in Fig. 4(c), where the TER peaks vary in both magnitude and position with QW thickness. In particular, the TER peak shifts toward more negative bias as the QW thickness increases, indicating that the resonant-voltage condition can be tuned through QW thickness. The corresponding RA characteristics in Fig. 4(d) also vary with applied voltage and QW thickness, further demonstrating the influence of QW thickness on the transport characteristics. These results establish the uniform double-QW structure as a reference for investigating thickness non-uniformity and show that QW thickness can tune the resonant transport even when the two QWs have identical dimensions. Accordingly, non-uniform QW thicknesses are investigated in the following section to further examine their effect on the relative alignment of the resonant states.

### C. Non-Uniform Thickness Engineering

Fig 5(a) and Fig 5(b) show the RA and TER, respectively, as functions of the QW thicknesses, $\mathrm{t_{qw1}}$ and $\mathrm{t_{qw2}}$, with fixed IL thicknesses of $\mathrm{t_{il1}}$=$\mathrm{t_{il2}}$=1 nm. The transport characteristics show a strong dependence on the QW thickness distribution. A favorable TER–RA combination is obtained for $\mathrm{t_{qw1}}$=1.9 nm and $\mathrm{t_{qw2}}$=3 nm, with a TER of $2.46 \times 10^7\%$ and an RA of $17.55\ \Omega \cdot \mathrm{cm}^2$. This QW thickness distribution is selected for further analysis of the resonant transport.

*1) Non-Uniform QW Thickness with Uniform IL Thickness:* The effect of QW thickness non-uniformity is first investigated with $\mathrm{t_{qw1}}$=1.9 nm and $\mathrm{t_{qw2}}$=3 nm, while keeping the IL thicknesses fixed at $\mathrm{t_{il1}}$=$\mathrm{t_{il2}}$=1 nm. The corresponding transport characteristics are presented in Fig. 6. The J-$\mathrm{V_a}$ characteristics in Fig. 6(a) show a clear LRS–HRS current difference in the low-bias region, while Fig. 6(b) shows a sharp TER peak near $\mathrm{V_r}$=12.5 mV together with a relatively low RA at the selected read voltage. To clarify the origin of this transport response, the corresponding energy-band profiles and transmission characteristics are examined in Fig. 6(c)–(f). In the LRS, two closely spaced resonant states are located at $\mathrm{E_1}$=0.282 eV and $\mathrm{E_2}$=0.288 eV at zero bias, giving an energy separation of only a few meV. These states produce two neighboring transmission peaks in Fig. 6(e), resulting in strong resonant transmission. In the HRS, the corresponding states occur at 0.453 eV and 0.627 eV, giving a much larger energy separation and a substantially weaker transmission response in Fig. 6(f). The polarization-dependent resonant-state alignment therefore provides the physical origin of the large LRS–HRS transmission difference. The next step is to examine how the QW thickness distribution controls the resonant-state alignment. This dependence is evaluated using $E_2$-$E_1$ in Fig. 6(g) and Fig. 6(h). For the LRS, $E_2$-$E_1$ decreases with increasing $t_{qw1}$, reaches a minimum near $t_{qw1}$=1.9 nm, and then increases. The selected $t_{qw1}$=1.9 nm therefore corresponds to a closely spaced resonant-state configuration in the LRS. The HRS exhibits a different evolution of $E_2$-$E_1$, resulting in a distinct resonant-state configuration. The insets further show the evolution of $E_1$ and $E_2$, confirming that QW thickness directly modifies the confined-state energies.

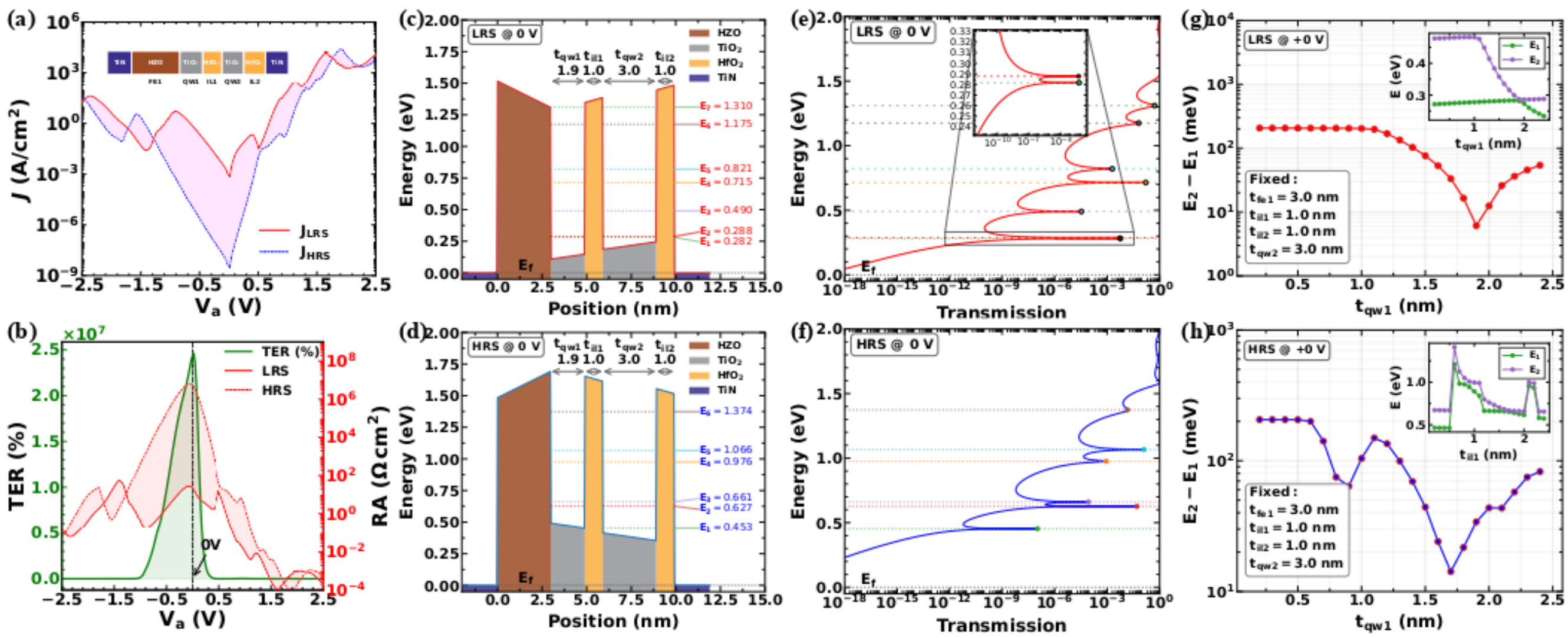

Fig. 6. Voltage-dependent transport and polarization-dependent resonant-state characteristics of the FTJ-(b3) structure with $\mathbf{t_{qw1}=1.9}$ nm, $\mathbf{t_{qw2}=3}$ nm, and $\mathbf{t_{il1}=t_{il2}=1}$ nm. (a) J-$\mathbf{V_a}$ characteristics under a voltage sweep of $\pm\mathbf{2.5}$ V and (b) corresponding TER and RA characteristics for the LRS and HRS. (c), (d) Energy-band profiles and (e), (f) corresponding transmission characteristics for the LRS and HRS, respectively. (g), (h) Resonant-state energy separation, $\mathbf{E_2}$-$\mathbf{E_1}$, as a function of $\mathbf{t_{qw1}}$ for the LRS and HRS, respectively, with $\mathbf{t_{qw2}=3}$ nm and $\mathbf{t_{il1}=t_{il2}=1}$ nm. The insets show the corresponding resonant-state energies.

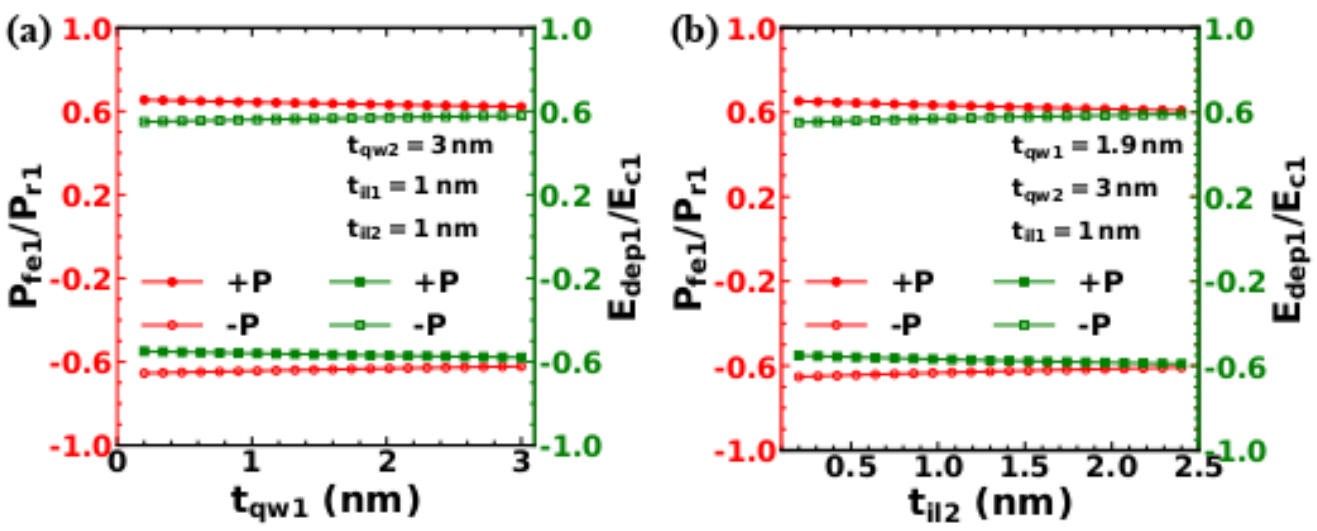

Fig. 7. Normalized polarization ($\mathbf{P_{fe1}/P_{r1}}$) and depolarization field ($\mathbf{E_{dep1}/E_{c1}}$) of the FTJ-(b3) structure evaluated at 0 V as functions of (a) $\mathbf{t_{qw1}}$ with fixed $\mathbf{t_{qw2}=3}$ nm, $\mathbf{t_{il1}=1}$ nm, and $\mathbf{t_{il2}=1}$ nm, and (b) $\mathbf{t_{il2}}$ with fixed $\mathbf{t_{qw1}=1.9}$ nm, $\mathbf{t_{w2}=3}$ nm, and $\mathbf{t_{il1}=1}$ nm.

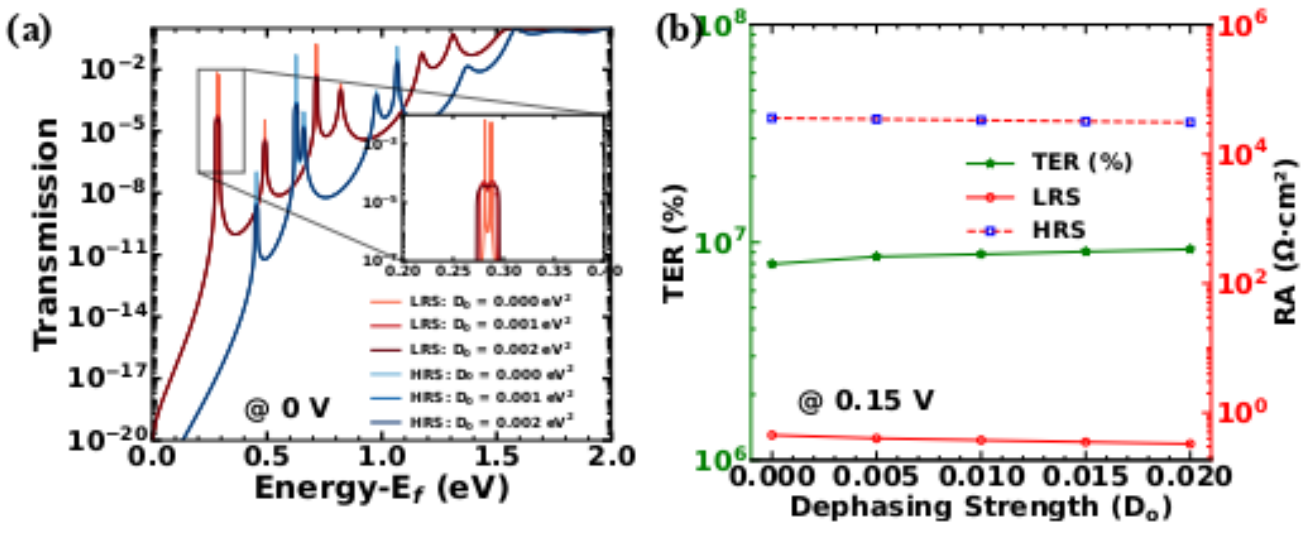

Fig. 8. Influence of elastic scattering on the transport characteristics of the FTJ-(b3) structure with $\mathbf{t_{qw1}=1.9}$ nm, $\mathbf{t_{qw2}=3}$ nm, $\mathbf{t_{il1}=1}$ nm, and $\mathbf{t_{il2}=1}$ nm. (a) Transmission spectra for various values of the dephasing strength, $\mathbf{D_0}$, at 0 V. (b) Variation of the TER and RA with the dephasing strength at $\mathbf{V_r}$=0.15 V.

The corresponding FE polarization and depolarization characteristics are examined in Fig. 7. In Fig. 7(a), $P_{fe1}/P_{r1}$ and $E_{dep1}/E_{c1}$ show only slight variations with $t_{qw1}$, with $t_{qw2}$ and the IL thicknesses fixed. Similarly, Fig. 7(b) shows only small changes in the normalized polarization and depolarization field with increasing $t_{il2}$, while the QW thicknesses and $t_{il1}$ are fixed. These relatively small variations indicate that the pronounced changes in the transport characteristics are mainly associated with thickness-dependent resonant transport rather than significant changes in the FE polarization state.

The influence of elastic scattering (momentum and phase) on the selected non-uniform QW structure is further examined in Fig. 8. The transmission spectra in Fig. 8(a) show distinct resonant peaks in the LRS at $D_0$=0. With increasing $D_0$, these resonances broaden and progressively overlap due to phase and momentum-randomizing scattering, leading to a smoother transmission response around the resonant-energy region. The HRS transmission remains substantially weaker over the investigated dephasing range. As shown in Fig. 8(b), the TER increases slightly, while the RA decreases with increasing $D_0$. These results show that elastic scattering by broadening the transmission peaks, can positively influence the performance of non-uniform multi-QW based FTJ in contrast to single QW based FTJs [17] and also maintain a clear difference between the LRS and HRS transport responses. With $t_{qw1}$=1.9 nm and $t_{qw2}$=3 nm fixed, the effect of IL thickness is investigated next to examine its additional influence on the resonant transport.

*2) Non-Uniform QW and IL Thicknesses:* With the QW thicknesses fixed at $t_{qw1}$=1.9 nm and $t_{qw2}$=3 nm, the IL thicknesses are varied to examine their additional influence on the resonant transport. Fig. 5(c) and Fig. 5(d) show the RA and TER, respectively, as functions of $t_{il1}$ and $t_{il2}$. The TER increases as $t_{il1}$ approaches 1.6 nm, but this enhancement is accompanied by a substantial increase in RA. At $t_{il1}$=1.4 nm and $t_{il2}$=2 nm, a high TER is maintained with a considerably lower RA. Therefore, this IL thickness distribution is selected to obtain a favorable TER–RA combination.

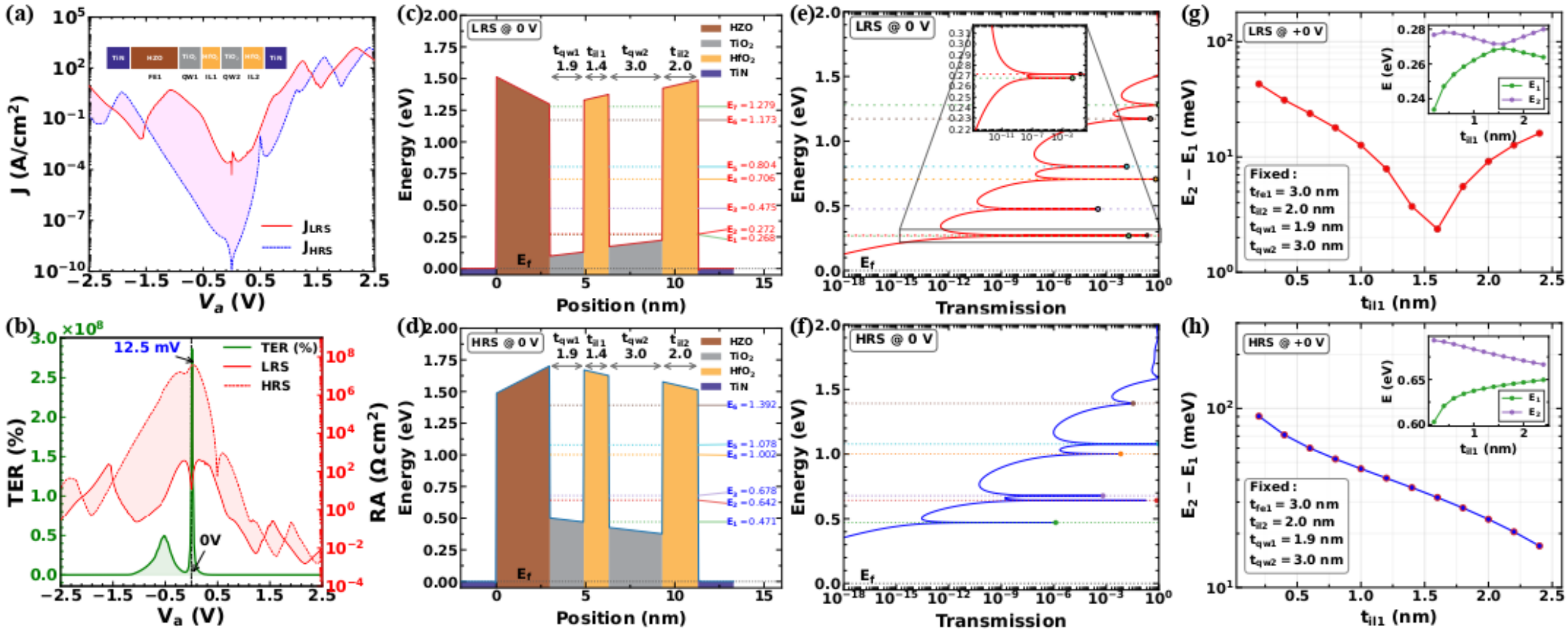

Fig. 9. Voltage-dependent transport and polarization-dependent resonant-state characteristics of the FTJ-(b3) structure with $\mathbf{t_{qw1}=1.9}$ nm, $\mathbf{t_{qw2}=3}$ nm, $\mathbf{t_{il1}=1.4}$ nm, and $\mathbf{t_{il2}=2}$ nm. (a) J-$\mathbf{V_a}$ characteristics under a voltage sweep of $\pm\mathbf{2.5}$ V and (b) corresponding TER and RA characteristics for the LRS and HRS. (c), (d) Energy-band profiles and (e), (f) corresponding transmission characteristics for the LRS and HRS, respectively. (g), (h) Resonant-state energy separation, $\mathbf{E_2}$-$\mathbf{E_1}$, as a function of $\mathbf{t_{il1}}$ for the LRS and HRS, respectively. The insets show the corresponding resonant-state energies.

The corresponding transport characteristics are presented in Fig. 9. The J-$V_a$ characteristics in Fig. 9(a) show a clear LRS–HRS current difference in the low-bias region. In Fig. 9(b), the TER exhibits a sharp maximum near $V_r$=12.5 mV together with a relatively low RA. At this read voltage, the selected structure achieves a TER of $2.39 \times 10^8\%$ and an RA of $24.3\ \Omega \cdot \mathrm{cm}^2$. The energy-band profiles and transmission spectra in Fig. 9(c)-(f) show strong resonant transmission in the LRS and a substantially weaker response in the HRS. In the LRS, the two resonant states are located at $E_1$=0.268 eV and $E_2$=0.272 eV, with an energy separation of 3.72 meV, resulting in strong resonant transmission in Fig. 9(e). In the HRS, the corresponding resonant states exhibit a much larger energy separation, leading to a substantially weaker transmission response in Fig. 9(f). The distinct resonant-state alignment between the LRS and HRS therefore accounts for the large difference in their transmission characteristics. The resonant-state separation, $E_2$-$E_1$, is shown as a function of $t_{il1}$ in Fig. 9(g) and Fig. 9(h) for the LRS and HRS, respectively. In the LRS, $E_2$-$E_1$ decreases with increasing $t_{il1}$, reaches a minimum near $t_{il1}$=1.6 nm, and then increases. The selected $t_{il1}$=1.4 nm lies close to this minimum, corresponding to closely spaced resonant states and strong LRS transmission. In the HRS, $E_2$-$E_1$ decreases steadily over the investigated range, resulting in a different resonant-state configuration. The corresponding insets of $E_1$ and $E_2$ as functions of $t_{il1}$ further demonstrate the role of IL thickness in tuning the resonant-state energies.

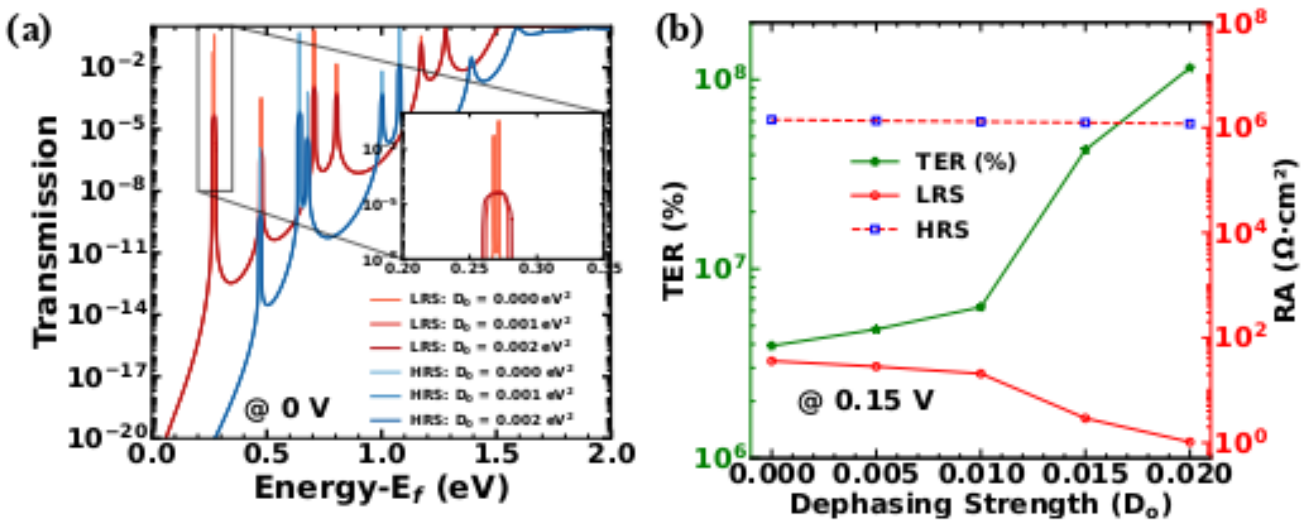

Fig. 10. Influence of elastic scattering on the transport characteristics of the FTJ-(b3) structure with $\mathbf{t_{qw1}=1.9}$ nm, $\mathbf{t_{qw2}=3}$ nm, $\mathbf{t_{il1}=1.4}$ nm, and $\mathbf{t_{il2}=2}$ nm. (a) Transmission spectra for different dephasing strengths, $\mathbf{D_0}$, at 0 V. (b) TER and RA as functions of $\mathbf{D_0}$ at $\mathbf{V_r}$=0.15 V.

The influence of elastic scattering on the selected non-uniform QW and IL structure is further examined in Fig. 10. Fig. 10(a) shows the transmission spectra for different dephasing strengths, $D_0$. At $D_0$=0, the LRS exhibits two closely spaced resonant peaks, as shown in the zoomed region, indicating strong resonant transmission within a narrow energy range. With increasing $D_0$, these resonances broaden and progressively overlap due to phase and momentum-randomizing scattering, resulting in a smoother transmission response around the resonant-energy region. The HRS transmission remains substantially weaker over the investigated range. As shown in Fig. 10(b), the TER increases with $D_0$, while the RA decreases significantly, particularly at higher dephasing strengths. This behavior is associated with the dephasing-induced broadening and overlap of the closely spaced LRS resonances, which modifies the LRS transmission while the HRS transmission remains relatively weak. Thus, elastic scattering modifies the resonant transmission and consequently positively influencing the TER–RA characteristics while maintaining a clear LRS–HRS transport difference.

## IV. Conclusion

In conclusion, non-uniform quantum well and barrier thickness engineering provides an approach for tailoring resonant transport in $HfO_2$-based multi-QW FTJs. The asymmetric thickness configuration modifies the alignment and spacing of resonant states, producing closely spaced resonances in the LRS and a larger separation in the HRS, which leads to strong polarization-dependent transmission. The self-consistent NEGF analysis further shows that elastic scattering broadens and increases the overlap of the closely spaced LRS resonances, thereby modifying the TER and RA characteristics. These results demonstrate that QW and barrier thickness can be used to control resonant-state alignment and transport characteristics in $HfO_2$-based multi-QW FTJs, providing a route toward high TER and low RA for high-density NVM applications.